# Kinesin Motor Transport is Altered by Macromolecular Crowding and Transiently Associated Microtubule-Associated Proteins

Leslie Conway,* Jennifer L. Ross,*[†]

*Department of Physics, University of Massachusetts Amherst, Amherst, MA USA; [†] Corresponding author

ABSTRACT   Intracellular transport of vesicular cargos, organelles, and other macromolecules is an essential process to move large items through a crowded, and inhomogeneous cellular environment. In an effort to dissect the fundamental effects of crowding and an increasingly complex cellular environment on the transport of individual motor proteins, we have performed in vitro reconstitution experiments with single kinesin-1 motors walking on microtubules in the presence of crowding agents and transient microtubule-associated proteins that more closely emulate the cellular environment. Macromolecular crowding due to inert polymers caused enhanced run lengths of motors, but displayed an increased tendency for non-specific motor association and diffusion, most likely due to depletion interactions. We found that transiently bound associated proteins slowed forward motion, but did not drastically affect the association times, in opposition to previously reported obstacle properties of stably associated microtubule-associated proteins, such as the neuronal protein tau. Such studies of the transport properties of molecular motors in increasingly complex reconstituted environments are important to illuminate the fundamental biophysical principles underlying the essential process of intracellular cargo transport.

Conventional kinesin-1 is an essential cargo motor for long-range transport. Kinesin-1 was discovered as the major enzymatic component responsible for long-distance transport along microtubules within the axon (1). Given this important role, it is not surprising that kinesin-1 mutants result in a variety of neurological diseases (2). Kinesin-1 motors typically work individually or in very small teams to transport large vesicular cargos around the cell (3), thus understanding how single kinesin-1 motors function has implications for intracellular cargos.

Kinesin motility is studied using single molecule biophysical techniques. Such studies have illuminated a number of important results about the mechano-enzymatic properties of kinesin-1 (4, 5). From these studies, we know that kinesin-1 has a tightly controlled and regular step-size of 8 nm, and can take hundreds of processive steps before dissociating from the microtubule. Such studies are performed using purified components without contaminating microtubule-associated proteins (MAPs) so that the characteristics of kinesin alone can be determined.

Using such clean, rigorous, single molecule studies as a basis, we systematically add back cellular components to increase the complexity of in vitro reconstitution to more closely model the cellular system (6). Several groups have examined kinesin motility in the presence of stabily-associated MAPs or stalled motors that can act as obstacles (7, 8). Other work has examined how the architecture of the roadways themselves can alter the motion of single motors (6). Such studies have shown that individual kinesin-1 motors are particularly susceptible to dissociation when the path is blocked (7, 8). Yet, all prior experiments have been missing two important aspects known to exist in the cell: macromolecular crowding and transient interactions of MAPs. In this study, we choose to focus on these two previously neglected aspects.

Macromolecular crowding has a number of significant effects on biochemical and biophysical processes. Crowding alters the available space and viscosity of a solution, increases the osmotic pressure or depletion interactions that force components together, enhancing chemical reactions (9). To test the effect of macromolecular crowding on kinesin-1 motility, we added in 5% 40 kD polyethylene glycol (PEG), a concentration above c*, the overlap concentration. When we add PEG to our single kinesin-1 experiments, we find the average run length and association time more than double compared to single kinesin motors moving along single microtubules in the absence of PEG (Fig. 1B, D), yet there is no effect on the velocity of motors (Fig. 3E).

Interestingly, the the association time for kinesin in the presence of PEG displays two characteristic times (see SI for details) that is not observed in the same-day control without PEG. The long decay is indicative of a second, slower process we believe is one-dimensional diffusion induced by depletion forces. In corroboration of this idea, we observed single kinesin motors that reversed direction on microtubules in the presence of PEG. We quantified the number of observable reversals in direction for single kinesin-1 motors in the presence and absence of PEG. We found that PEG causes reversals of about 10% of single kinesin motors, whereas zero kinesin-1 motors reverse in the absence of PEG (Fig. 2A).

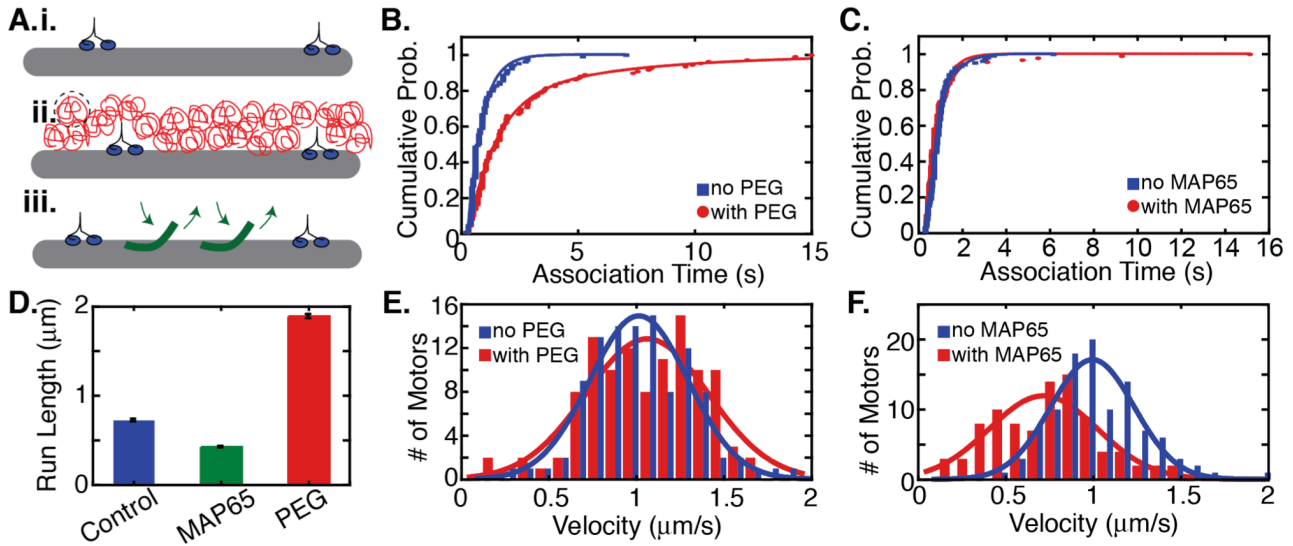

**Figure 1.** Kinesin motility affected by PEG or MAP65. **A.** Cartoons of experimental systems: **(i)** control, **(ii)** with PEG (red random coils), or **(iii)** with MAP65 (green transient MAPs). **B.** Association time of kinesin in the presence of PEG has two characteristic decay constants (red circles and red fit line), compared to control (blue squares and blue fit line). **C.** Association time of kinesin in the presence of MAP65 (red circles, red line fit) is the same as control without MAP65 (blue squares, blue line fit). **D.** The characteristic decay times for the run length are shorter in the presence of MAP65 (green bar) and longer in the presence of PEG (red bar) compared to control (blue bars). **E.** Velocity of kinesin in the presence of PEG (red bars) is the same as control (blue bars). **F.** Velocity of kinesin in the presence of MAP65 (red bars) is slower compared to control (blue bars).

Reversals are quite unheard of for kinesin stepping, and back-stepping has only be achieved at high load (10). Closer examiniation of the traces gave the impression that kinesin-1 motors enter a diffusive state while bound to microtubules (Fig. 2B, C). To quantify this, we calculated the mean-squared displacement (MSD) and the frame-to-frame displacement for individual motors that appeared to reverse (Fig. 2B, C, and Fig. S1). Three of the 8 reversing traces showed MSDs that were dominated by ballistic motion and showed parabolic traces (Fig. 2B, ii). For these traces, the frame-to-frame displacements showed significantly more motion in the plus-end direction compared to the minus-end direction, indicating the overall ballistic motion (Fig. 2Biii, Fig. S1).

Other motors that reversed showed MSDs dominated by diffusive motion and showed no ballistic-like behavior (Fig. 2C, Fig. S1). The frame-to-frame displacements showed an almost even distribution toward the minus-end and plus-end as expected for a random, diffusive walk (Fig. 2Ciii, Fig. S1). Such diffusive motion implies that the kinesin motors are not bound to the microtubule with the ATP-dependent microtubule binding site, but rather are non-specifically bound.

The depletion forces caused by the presence of the PEG molecules likely enhances the non-specific binding of motors to the microtubule. This has two effects: first, it enhances the association time and run length of single motors, and second, it causes some of the molecules to associate to the microtubules non-specifically and diffuse along the microtubule.

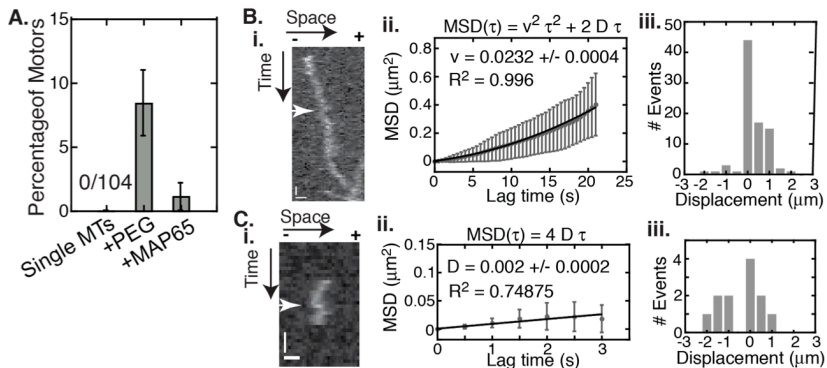

**Figure 2.** PEG induces diffusion of kinesin. **A.** Kinesin appears to reverse in the presence of PEG compared to control. **B. i.** Kymograph of single kinesin in the presence of PEG displays reversing (arrow head). Microtubule plus-end on the right side; distance horizontal direction; time positive in the downward direction. **ii.** The mean-squared displacement with significant diffusive and ballistic components. **iii.** The number of frame-to-frame displacements show this motor moved mostly toward the microtubule plus-end. **C. i.** Kymograph of single kinesin in the presence of PEG displays reversing (arrow head). **ii.** The mean-squared displacement had significant diffusive component and no clear ballistic component. **iii.** The frame-to-frame displacements shows almost equal displacements toward the minus-end and plus-end of the microtubule, indicating diffusion.

In addition to macromolecular crowding, in cells, many protein complexes are transiently associated, not long-lived nor stable (11). Prior work testing the effects of MAPs on motor protein activity have used stably associated proteins, such as the neuronal MAP, tau, that block kinesin binding sites and act as obstacles to kinesin motion. This is shown to result in the premature dissociation of motors, causing motors to have reduced association times and run lengths, while having no effect on the velocity of the motor (7, 8). Although certain MAPs may serve as stable roadblocks, recent work has shown that numerous MAPs are not stable, but rather transiently bound to microtubules in cells and in vitro (12, 13). Even tau may also be transiently bound to microtubules in vivo and in some in vitro experiments (14).

To investigate the effects of a transiently bound MAP on the activity of kinesin motility, we used the plant protein, MAP65 that binds with a short association time (12). We added 112 nM MAP65 to the sample with kinesin motors. Unlike PEG, MAP65 does not change the association time of single kinesin motors to the microtubule (Fig. 1C). However, we observe that MAP65 causes a reduction in both the run length (Fig. 1D) and velocity of kinesin motors (Fig. 1F) implying that kinesin-1 takes fewer steps in the presence of MAP65, but the time of each step must be slightly longer. This is in agreeement with the slower motor velocity measured. Future studies measuring the ATP dependence of this effect, and resolvling individual steps of kinesin motors in the presence of transient obstances would be interesting.

Our results highlight that physiological crowding and transient interactions of MAPs can have effects on kinesin motor transport. Interestingly, crowding increases the association time, but motors spend more time in diffusive states. Diffusion of kinesin motors is likely important for motor recycling in the axon (15), targeting motors using an antennae mechanism (16), or transport by monomeric kinesin motors (17). Transiently associated MAPs slow motors and reduce the number of steps they take before dissociating, but have fewer detrimental effects on motor motion than stable MAPs or other permanent obstacles. Our studies illuminate two novel and physically-based motor regulating mechanisms that are likely to affect the transport of motors in living cells.

**SUPPORTING MATERIAL:**

Materials and Methods and two figures are available online.

**ACKNOWLEDGEMENTS:**

We thank Dr. Michael Gramlich for helpful discussions. This work was supported by NSF-DMR #1207783 to JLR, a Mathers Foundation grant to JLR, and a Cottell Scholars Grant #20031 from Research Corporationfor Science Advancement. LC was partially supported an NSF-IGERT DGE#0654128 to Susan Roberts.

# Supporting Information

# Kinesin Motor Transport is Altered by Macromolecular Crowding and Transiently Associated Microtubule-Associated Proteins

Leslie Conway,* Jennifer L. Ross,*[†]
*Department of Physics, University of Massachusetts Amherst, Amherst, MA USA; [†] Corresponding author

**Single Molecule Association Time and Run Length Data and Fits:**
   The process of motor stepping is terminated by dissociation of the motor. Kinesin-1 motors step repeatedly, processively for many steps. At any step, there is some fixed probability that the motor will dissociate. Such processes result in exponentially decaying probability distributions. In order to observe the exponential decay, the data must be binned and plotted as a histogram. Since there is a minimal exposure time of 0.5 s for our data, we are unable to observe associations that are shorter on the order of 0.5 s, and the lowest time bin is undercounted. Further, decisions as to the best bin size are arbitrary, and can result in different interpretations of the data. In order to avoid these issues, we choose to plot the data as a cumulative probability distribution. We did this by rearranging the data in ascending order (lowest to highest association times). Then we created a list from 1 to N for the number of measurements made. In order to normalize the distribution, the list was divided by N, so that the largest association time reached the maximum probability of 1. We plotted the ascending association time data on the x-axis and the list from 1/N to 1 on the y-axis. In this form, the data exponentially approach 1 instead of zero, and can be fit to the functional form:

$$CDF = 1 - A\exp(-t/\tau_c) \quad (1)$$

where CDF is the cumulative distribution function, A is the amplitude, t is the association time plotted on the x-axis as the independent axis, and $\tau_C$, is the characteristic decay time. We used this functional form to fit the data for the association time of the control data and the association time of the MAP65 data (Fig S1, A, B). The data for single kinesin motors in the presence of PEG did not fit well to the single exponential form, but fit well to a double exponential of the form:

$$CDF = 1 - A\exp(-t/\tau_{short}) - B\exp(-t/\tau_{long}) \quad (2)$$

where $\tau_{short}$ is the shorter characteristic decay time and $\tau_{long}$ is the longer characteristic decay time.; B is the amplitude of the second decay (Fig. S1, B). As shown by the residuals (Fig. S1, A, inset) the single exponential does not fit well to the long time data and has systematic errors. The residuals of the double exponential decay are scattered about zero, and are smaller than for the single exponential decay fit.
   The run length for the control data, PEG data, and MAP65 data also fit well to the single exponential functional form (Eq. 1), with the change that the run length of the motors was plotted on the x-axis and a characteristic run length, $x_C$, was found (Fig. S1, C, D). The data plotted in Figure 1, D is the characteristic run length found by fitting the data to the functional form.

**Single Molecule Velocity Data and Fits:**
   The velocity of single molecule kinesin-1 motors in control experiments and in the presence of PEG or MAP65 were binned, plotted as a histogram, and fit with Gaussian functions of the form:

$$PDF = A\exp\left(-\frac{(x-x_0)^2}{2\sigma^2}\right) \quad (3)$$

where PDF is the probability distribution function, A is the amplitude, $x_0$, is the mean of the distribution, σ is the standard deviation of the distribution. We chose to use the probability distribution instead of the cumulative distribution because the bin size did not affect the data: the mean and standard deviation were insensitive to bin size. Further, the fit for the CDF for a normal data set is an error function, which our fitting program, KaleidaGraph, is unable to perform. The best fit parameters are given (Fig. S1, E, F).

**Mean Squared Displacement Analysis and Fits:**
The trajectory of single kinesin motors that appeared to reverse were analyzed using the mean-squared displacement (MSD), and a frame-to-frame displacement analysis. The MSD was computed by taking the difference in the displacement over a lag time, $\tau$. These data sets were best fit to a quadratic equation to determine how the motion depends on time (diffusive-like) and time squared (processive-like) using this functional form:

$$MSD(\tau) = v^2\tau^2 + 2D\tau \qquad (4)$$

where $\tau$ is the lag time, v is the velocity, and D is the diffusion coefficient. For these same traces, the frame-to-frame displacements showed significantly more motion in the plus-end direction compared to the minus-end direction again indicating the overall ballistic motion (Fig. 2Biii, Fig. S1).

Other motors that reversed showed MSDs that were dominated by diffusive motion and showed no ballistic-like behavior that depended on the lag time squared (Fig. 2C). For these traces, we fit the MSD to a diffusive equation:

$$MSD(\tau) = 4D\tau \qquad (5)$$

which is linear in the lag time, $\tau$. Five of the 8 traces were best fit by the linear equation (Fig. 2Cii, Fig. S1).

Supplemental Figure Captions:

**Figure S1**: **Distribution of association times, run lengths, and velocities for kinesin motors in the presence of PEG or MAP65 with same-day controls in the absence of PEG or MAP65 with fits and fit parameters.**
**A.** Association times of kinesin in the presence of PEG (red circles) or in the absence of PEG (blue squares) have distinct characteristics. Control data in the absence of PEG is a single exponential decay fit (blue fit line) to equation (1) with fit parameters given (blue font). In the presence of PEG, the association times are distributed biexponentially, and fit (red fit line) to equation (2) with fit parameters given (red font).
**(Inset)** To corroborate that the biexponential is the best fit, we plot the residuals of the data with PEG with a single exponential fit (green circles) or a biexponential fit (purple squares). The residuals for the single exponential fit systematically miss the experimental data resulting in large residuals, especially at long times. The double exponential decay fit has smaller and more random residuals.
**B.** Association times of kinesin in the presence of MAP65 (red circles) or in the absence of MAP65 (blue squares) are identical. Both data sets are fit with a single exponential (Eq 1), and the best fit parameters are given for the control (blue fit line, blue font) and MAP65 experiment (red fit line, red font).
**C.** Run lengths of kinesin in the presence of PEG (red circles) or in the absence of PEG (blue squares) have distinct distributions. Both data sets can be fit with a single exponential (Eq 1), and the fit parameters of the best fit line for the control (blue line, blue font) and the data with PEG (red fit line, red font) are given. The characteristic decay length, $\tau_C$, is significantly shorter for the control data compared to the data with PEG. These characteristic decay lengths are reported in Fig. 1D.
**D.** Run lengths of kinesin in the presence of MAP65 (red circles) or in the absence of MAP65 (blue squares) have distinct distributions. Both data sets can be fit with a single exponential (Eq 1), and the fit parameters of the best fit line for the control (blue line, blue font) and the data with MAP65 (red fit line, red font) are given. The characteristic decay length, $\tau_C$, is significantly shorter for the data with MAP65 compared to the control data. These characteristic decay lengths are reported in Fig. 1D.
**E.** Velocities of kinesin in the presence of PEG (red bars) compared to the control without PEG (blue bars) have similar distributions. The data are fit using equation 3, and the parameters for the best fit lines are reported for control (blue fit line, blue font) and for the PEG data (red fit line, red font). There is very little difference in the mean value or the width of the distributions.

**F.** Velocities of kinesin in the presence of MAP65 (red bars) compared to the control without MAP65 (blue bars) have distinct distributions. The parameters for the best fit lines are reported for control (blue fit line, blue font) and for the MAP65 data (red fit line, red font). The kinesin motors move slower in the presence of MAP65, on average.

**Figure S2. Mean Squared Displacement and Step Direction of Motors on Single Microtubules in the Presence of PEG.**
**A-F.** For each example of a single kinesin that appears to reverse direction, we show:
**i**. the kymograph of the motion,
**ii.** a plot of the mean squared displacement (MSD) against the lag time, and
**iii.** a distribution of the frame-to-frame displacement for each trajectory.
For all kymographs, the microtubule plus end is oriented towards the right. Horizontal scale bars are 0.5 µm and vertical scale bars are 3 sec. For MSDs, data was plotted with an upper bound at half the total time of the trajectory so that the data represents the average of multiple data points, decreasing the uncertainty in the data. For each data point shown, at least N points were averaged. For A, N= 13; for B, N= 46; for C, N = 6; for D, N = 23; for E, N = 10; for F, N = 39. The frame-to-frame displacement for each trajectory is plotted such the positive displacements are in the plus-end direction, negative displacements are in the minus-end direction, and pauses have zero displacement.

**Association Times**   CPD = 1 - Aexp(-t/$\tau_C$)

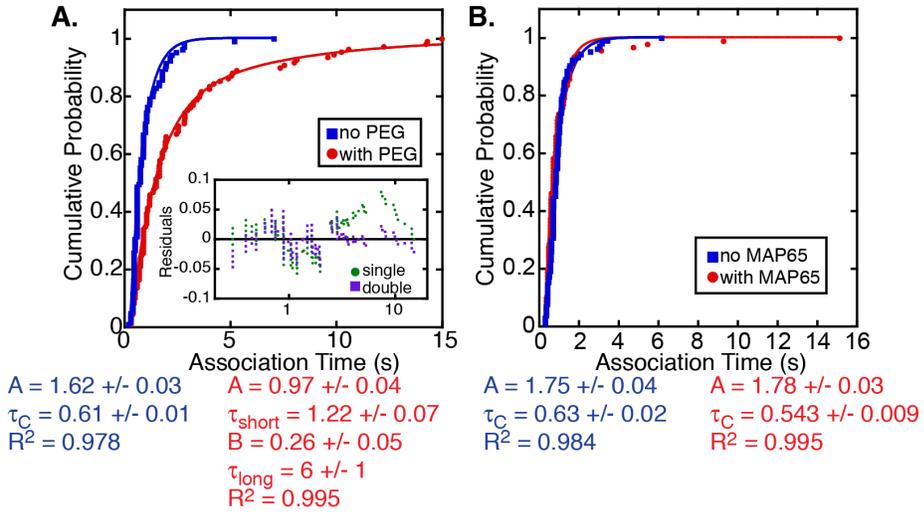

A.
A = 1.62 +/- 0.03
$\tau_C$ = 0.61 +/- 0.01
$R^2$ = 0.978

A = 0.97 +/- 0.04
$\tau_{short}$ = 1.22 +/- 0.07
B = 0.26 +/- 0.05
$\tau_{long}$ = 6 +/- 1
$R^2$ = 0.995

B.
A = 1.75 +/- 0.04
$\tau_C$ = 0.63 +/- 0.02
$R^2$ = 0.984

A = 1.78 +/- 0.03
$\tau_C$ = 0.543 +/- 0.009
$R^2$ = 0.995

**Run Lengths**   CPD = 1 - Aexp(-t/$\tau_C$)

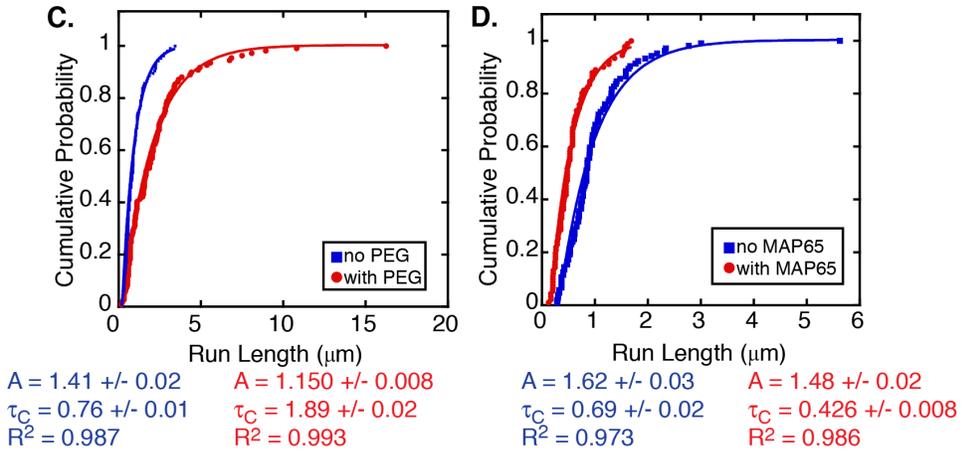

C.
A = 1.41 +/- 0.02
$\tau_C$ = 0.76 +/- 0.01
$R^2$ = 0.987

A = 1.150 +/- 0.008
$\tau_C$ = 1.89 +/- 0.02
$R^2$ = 0.993

D.
A = 1.62 +/- 0.03
$\tau_C$ = 0.69 +/- 0.02
$R^2$ = 0.973

A = 1.48 +/- 0.02
$\tau_C$ = 0.426 +/- 0.008
$R^2$ = 0.986

**Velocities**   PDF = Aexp(-(x-x0)$^2$/2$\sigma^2$)

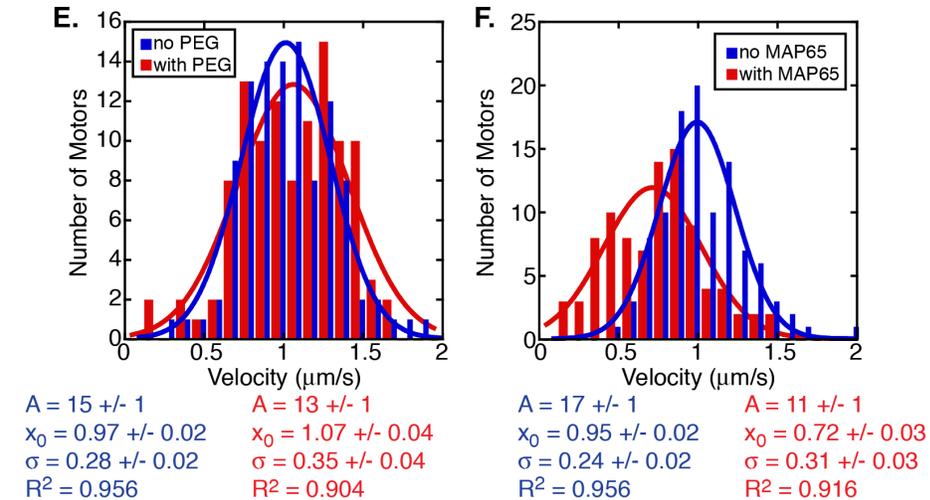

E.
A = 15 +/- 1
$x_0$ = 0.97 +/- 0.02
$\sigma$ = 0.28 +/- 0.02
$R^2$ = 0.956

A = 13 +/- 1
$x_0$ = 1.07 +/- 0.04
$\sigma$ = 0.35 +/- 0.04
$R^2$ = 0.904

F.
A = 17 +/- 1
$x_0$ = 0.95 +/- 0.02
$\sigma$ = 0.24 +/- 0.02
$R^2$ = 0.956

A = 11 +/- 1
$x_0$ = 0.72 +/- 0.03
$\sigma$ = 0.31 +/- 0.03
$R^2$ = 0.916

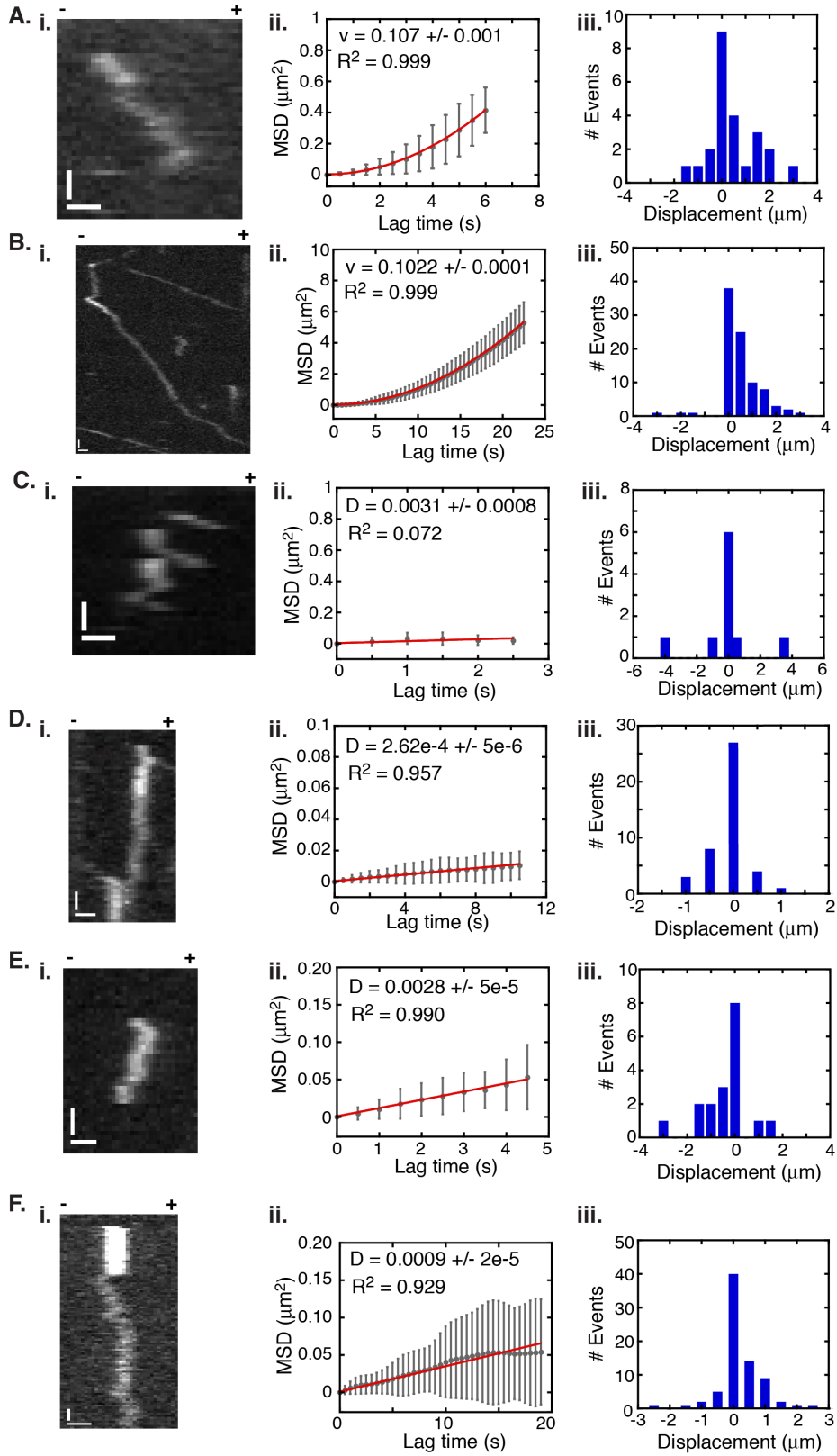